\documentclass{aa}

\newcommand{\nuc}[3]{$^{#1#2}$#3}

\usepackage{graphicx}
\usepackage{txfonts}
\usepackage{natbib}
\bibpunct{(}{)}{;}{a}{}{,} 

\begin{document}

   \title{Uncertainties and robustness of the ignition process in type Ia supernovae}
   \titlerunning{Uncertainties and robustness of the ignition in SNe Ia}

   \author{L. Iapichino\inst{1}
          \and
          P. Lesaffre\inst{2,3}
          }

   	      \institute{Zentrum f\"ur Astronomie der Universit\"at 
	      Heidelberg, Institut f\"ur Theoretische Astrophysik, 
	      Albert-Ueberle-Strasse 2, D-69120 Heidelberg, Germany \\
              \email{luigi@ita.uni-heidelberg.de}
              \and
              Laboratoire de Radioastronomie, \'Ecole Normale Sup\'erieure,
              24 rue Lhomond, 75231 Paris Cedex 05, France \\
              \email{pierre.lesaffre@lra.ens.fr}
	      \and
	      Institute of Astronomy, Madingley Road, Cambridge CB30HA, UK
              }
              
\date{Received 17 September 2009/ Accepted 15 December 2009}

   \abstract{It is widely accepted that the onset of the explosive carbon burning in the core of a carbon-oxygen white dwarf (CO WD) triggers the ignition of a type Ia supernova (SN Ia). The features of the ignition are among the few free parameters of the SN Ia explosion theory.}{We explore the role for the ignition process of two different issues: firstly, the ignition is studied in WD models coming from different accretion histories. Secondly, we estimate how a different reaction rate for C-burning can affect the ignition.}{Two-dimensional hydrodynamical simulations of temperature perturbations in the WD core (``bubbles'') are performed with the {\small FLASH} code. In order to evaluate the impact of the C-burning reaction rate on the WD model,  the evolution code {\small FLASH\_THE\_TORTOISE} from Lesaffre et al.~(2006) is used.}{In different WD models a key role is played by the different gravitational acceleration in the progenitor's core. As a consequence, the ignition is disfavored at a large distance from the WD center in models with a larger central density, resulting from the evolution of initially more massive progenitors. Changes in the C reaction rate at $T \lesssim 5 \times 10^8\ \mathrm{K}$ slightly influence the ignition density in the WD core, while the ignition temperature is almost unaffected. Recent measurements of new resonances in the C-burning reaction rate (Spillane et al.~2007) do not affect the core conditions of the WD significantly.}{This simple analysis, performed on the features of the temperature perturbations in the WD core, should be extended in the framework of the state-of-the-art numerical tools for studying the turbulent convection and ignition in the WD core. Future measurements of the C-burning reactions cross section at low energy, though certainly useful, are not expected to affect dramatically our current understanding of the ignition process.}

\keywords{supernovae: general -- hydrodynamics -- methods: numerical -- nuclear reactions, nucleosynthesis, abundances -- white dwarfs}

\maketitle

\section{Introduction}
\label{intro}

The sequence of events that leads to the ignition of thermonuclear flames in the core of mass-accreting carbon-oxygen white dwarfs (CO WD) marks the start of the explosion of a type Ia supernova (SN Ia). It is known from theory \citep{hn00} that the ignition process spans over a broad range of typical length and time scales, corresponding to the transition from hydrostatic to explosive C-burning. As a consequence, a proper modeling of this evolutionary phase is very challenging both from the theoretical and computational point of view. In the framework of the single degenerate explosion scenario, a detailed knowledge of the features of the WD at ignition is crucial for determining the initial conditions of the explosion models, i.e.~for shaping the initial flame morphology in the numerical simulations. 

The dependence of the explosion outcome and features on the initial spatial distribution of the flame kernels has been often debated, given the underlying relative freedom in constraining the unknowns of the multi-spot ignition scenario \citep{gsb05,rhn06,rwh07}. In general, it turns out that a larger number of ignition points results in a more vigorous deflagration phase. Interestingly, in recent models with deflagration to detonation transition (DDT), ignition closer to the center leads to a weaker detonation phase and a smaller production of {\nuc 56 Ni}, because of the stronger WD pre-expansion in the deflagration stage \citep{wkm08,krw09}.

Several different approaches have been used to address the problem of SN Ia ignition. Most of the work on this topic has been devoted to full-star simulations of the CO WD, both using analytical models \citep{wwk04,ww04} and numerical simulations \citep{hs02,kwg06,pb08,pc08,zab09}. A parallel branch of studies focuses on the role of the convective Urca process for the ignition (\citealt{lpt05,lpt05b}, and references therein). 

Recently \citet{lht06} studied the link between the evolution of the progenitor in its binary system and the ignition conditions of SNe Ia. They found a broad range of densities and temperatures for WD models at ignition, linked to the initial WD mass and the binary accretion history. The effect of the progenitor diversity on the explosion phase still needs to be explored. In particular, a fundamental issue has to be fully addressed, namely the relation between the properties of the ignition process and the homogeneity  of SNe Ia as a class of objects. Does ignition contribute to the homogeneity of SNe Ia or, on the contrary (cf.~\citealt{rgr06}), is it more intrinsically linked to the residual diversity of Branch-normal SNe Ia?  

A complementary contribution to the study of this problem is provided by the small-scale (1 km or less) modelling approach. These models are motivated by analytical calculations \citep{gsw95,ww04} which predict that temperature perturbations (``bubbles''), produced in the WD's core by the turbulent convective flow, finally trigger the thermonuclear runaway. The physics of the thermal fluctuation has been addressed by \citet{gsb05} and \citet{zd07}, which focus on the transition from hydrostatic to explosive burning and the propagation of the first thermonuclear flames. 
\citet{ibh06} studied the bubble physics with a series of small-scale 2D simulations. According to their results, the main feature of the bubble evolution is the fragmentation caused by its own buoyant rise motion, which can be in a sense considered as a cooling effect, counteracting the heating from the hydrostatic nuclear burning. Among nuclear burning and hydrodynamical dispersion, the process with the smaller timescale determines the outcome of the bubble, namely if it will trigger a thermonuclear runaway or will be dispersed and cooled down by thermal conduction.

The goal of the small-scale studies is the description of the SNe Ia ignition scenario in terms of number of flame kernels, size of the ignition region in the WD core, (a)symmetry of the ignition and stochastic behaviors of the process. There is general agreement that the WD convective flow prior to the thermonuclear runaway plays a crucial role on the ignition. It has also been proposed that the flame kernels could tend to cluster spatially, thus evolving in a sort of ignition plume, rather than separate flame seeds (e.g., \citealt{lah05}). This idea is closely linked to the flow in the WD convective core, which is likely to develop a dipole feature \citep{wwk04,ww04,kwg06,zab09}.

The study of \citet{ibh06} explored only the ignition parameters that are inherent  to the bubble physics: the initial bubble diameter, the initial bubble temperature, the initial distance of the bubble from the WD center. Though instructive for understanding the small-scale physics of the temperature perturbations, that work could only very generically constrain the large-scale ignition features, through a number of additional theoretical assumptions. In the present work, we widen the scope to further ignition parameters, which can be related to the ignition condition in a more straightforward way. In particular, the bubble simulations developed in \citet{ibh06} will be used to probe the ignition conditions in the WD models presented in \citet{lht06}. We will show how the large range of ignition densities in the aforementioned WD models can be linked to a significant diversity for the size of the ignition region.

In the second part of the present work, these results will be applied to a largely unaccounted issue of the ignition process, namely the assessment of the role of experimental uncertainties in the C-burning reaction rates. In a recent study, \citet{srr07} present new reaction rates for the reactions {\nuc 12 C}({\nuc 12 C}, $\alpha$){\nuc 20 Ne} and {\nuc 12 C}({\nuc 12 C}, p){\nuc 23 Na}, which are larger than previously assumed \citep{cf88}. Those authors suggest that the new measurements could affect the ignition and even the explosion theory of SNe Ia. We show in this work that the impact of these measurements on the WD evolution to ignition is actually marginal, and by using the  WD evolution code by \citet{lht06} we make estimates on the role of a hypothetical larger (or lower) rate. 

The structure of the paper is the following: in Sect.~\ref{tools} the simulations will be introduced, with special emphasis on the details related to the WD background models (Sect.~\ref{wd}) and the C-burning reaction network (Sect.~\ref{burning}). The result are presented in Sect.~\ref{results}, and the conclusions are drawn in Sect.~\ref{conclusions}.

\section{Numerical tools}
\label{tools}

\subsection{Bubble setup}
\label{setup}

The 2D numerical simulations presented in this work have been performed with the {\small FLASH} code (v.~2.3; \citealt{for00}), an hydrodynamical, adaptive mesh refinement (AMR) code especially designed for the study of thermonuclear flashes. The simulation setup is very similar to the one described in \citet{ibh06}, and we refer the reader to that work for a more detailed account. The most relevant differences with respect to it are presented in the next sections. A 2D bubble is initialized as a temperature perturbation, in pressure equilibrium with a background medium, in a computational domain with size $5 \times 20\ \mathrm{km}$. The simulations have Cartesian geometry and were set with a root grid of $[16 \times 64]$ grid cells and four additional AMR levels, leading to an effective grid size of $[256 \times 1024]$, and to an effective spatial resolution of $2 \times 10^3\ \mathrm{cm}$.  The initial bubble diameter is $D = 1\ \mathrm{km}$, considered to be a good order-of-magnitude estimate of the size of the temperature perturbations in the turbulent WD convective flow \citep{wwk04}. The initial distance of the bubble from the WD center is set to $R = 100\ \mathrm{km}$, and the initial temperature contrast $\Delta T / T$ is set to 0.07 at the bubble location. Both values are in the range predicted by \citet{ww04} for the generation of temperature fluctuations in the WD core, though these authors refer to the contrast to the core temperature rather than to the local temperature. Note that, in the framework of our simplified model, $R$ and $\Delta T / T$ are not completely independent parameters: taking a slightly bigger contrast 
at a given R is like considering a bubble that has already travelled and 
grown from a slightly smaller distance to the center. Therefore, as long as $\Delta T / T$ is relatively small, its value is not a critical issue in the model.

\subsection{WD models}
\label{wd}

Unlike \citet{ibh06}, we adopt here the WD models from \citet{lht06} as background for the simulations. The models are labelled in Table \ref{models} with the number $m$ in the first column, where $m=1,3,4,6$ is related to the initial mass of the WD: $M_{\mathrm{init}} = (0.6 + 0.1 \times m)\ M_{\odot}$. All models are for a cooling age of $t_a=0.4$~Gyr \citep[see][]{lht06}. In Table \ref{models} the quantity $T_\mathrm{c}$ is the central temperature of the model, $\rho_\mathrm{c}$ is the central density and $g$ is the gravitational acceleration at a distance of 103\ km from the WD center. The burning zone radius of the model $r_{\mathrm{b}}$ is defined as the radius within which one-half of the luminosity is generated.

\begin{table}
\caption{Main features of the four WD models used in the simulations.} 
\centering
\begin{tabular}{ccccc}
\hline \hline
Model  & T$_\mathrm{c}$ [$10^8\ \mathrm{K}$] & $\rho_\mathrm{c}$ [$10^9\ \mathrm{g\ cm^{-3}}$] & $g$ [$10^9\ \mathrm{cm\ s^{-2}}$] & $r_{\mathrm{b}}$ [km]\\
\hline
1 & 7.84 & 1.82 & 5.18 & 163\\
3 & 7.63 & 2.08 & 5.90 & 154\\
4 & 7.15 & 3.12 & 8.78 & 135\\
6 & 6.94 & 4.17 & 11.7 & 123\\
\hline
\end{tabular}
\label{models}
\end{table}

The details of the modeling of the accretion phase are reported in \citet{lht06}. The models are taken at the time in the accretion history when the burning time scale is equal to the convective turnover time scale (this corresponds to the ignition criterion $\alpha=1$ of \citealt{lht06}, Sect.~2.5).  The chemical composition of the WD is not set as in \citet{ibh06} (X($^{12}$C) = X($^{16}$O) = 0.5), but is mapped from the WD model. We pick the models from \citet{lht06} which use values from \citet{uny99}  as initial conditions before the accretion stage. The C/O ratio in the WD core at ignition ends up between 0.74 and 0.46, depending on the model.

The gravitational acceleration $g(r)$ is derived from the WD mass profile as

\begin{equation}
\label{accel}
g(r) = \frac{G M(r)}{r^2}
\end{equation}
where $G$ is the gravitational constant and $M(r)$ the mass contained in the spherical shell with radius $r$. The acceleration is applied to the computational domain as a force pointing downwards along the $y$-axis, constant in space and time.

\subsection{Nuclear burning and carbon reaction rate}
\label{burning}

In the initial WD models the abundances of four chemical species are mapped, i.e.~{\nuc 12 C}, {\nuc 16 O}, {\nuc 22 Ne} and {\nuc 24 Mg}. As in \citet{ibh06}, the hydrostatic nuclear burning is followed with the reaction network {\tt iso7} \citep{thw00} which evolves seven $\alpha$-isotopes. As {\nuc 22 Ne} is not available in the reaction network, this isotope is substituted by {\nuc 20 Ne}, as suggested and verified by \citet{dt06}.
We refer to \citet{thw00} for the nuclear rates in the reaction network, with the notable exception of the C-burning, which in some calculations is modified according to the following prescriptions. 

\citet{srr07} performed a study of the carbon burning reactions {\nuc 12 C}({\nuc 12 C}, $\alpha$){\nuc 20 Ne} and {\nuc 12 C}({\nuc 12 C}, p){\nuc 23 Na} using $\gamma$-ray spectroscopy with a C target with ultra-low hydrogen contamination. The explored energy range spanned from $2.10$ to $4.75\ \mathrm{MeV}$. The most relevant result is the detection of a new resonance at the very low end of the energy range ($E_{R} = 2.14\ \mathrm{MeV}$).

\begin{figure}
  \resizebox{\hsize}{!}{\includegraphics[angle=90]{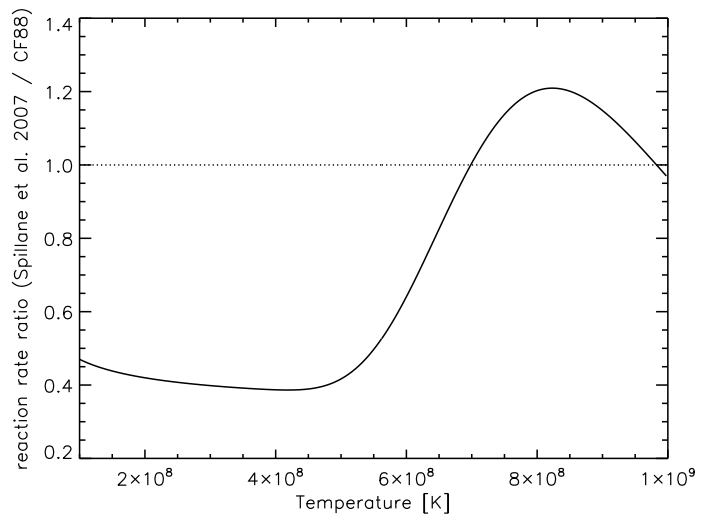}}
  \caption{Ratio (solid line) between the new reaction rate of \citet{srr07} for C-burning, and the \citet{cf88} rate, in a temperature range relevant for the ignition of SNe Ia. The non-resonant behavior at low temperature has been extrapolated from the analytic fit of \citet{srr07}, which refer to a higher energy range.}
  \label{rate-ratio}
\end{figure}

The effect of this new measurement for the C-burning reaction rate (the sum of the contributions of the $\alpha$ and proton channels) as a function of the temperature is shown in Fig.~\ref{rate-ratio}. With respect to the reference value of \citet{cf88}, the reaction rate is increased up to 20\% in the temperature range between $7.0$ and $9.8\ \times 10^8\ \mathrm{K}$. For comparison, the Gamow energy for the C-burning reactions is $1.5 \pm 0.3\ \mathrm{MeV}$ at $T = 5 \times 10^8\ \mathrm{K}$, and $2.1 \pm 0.4\ \mathrm{MeV}$ at $T = 8 \times 10^8\ \mathrm{K}$. The new resonance therefore affects the reaction rate mostly in the late hydrostatic burning stage, just prior to ignition. The data do not provide information about the burning rate at lower temperatures, relevant for the ``smouldering phase'' of the burning in massive CO WDs ($5 \times 10^7\ \mathrm{K} \lesssim T \lesssim 7 \times 10^8\ \mathrm{K}$; cf.~\citealt{n82}). An extrapolation at these temperatures rests upon the new evaluation of the non-resonant contribution to the reaction rate from \citet{srr07}, plus general considerations on the unknown resonance structure below $2.0\ \mathrm{MeV}$. Assuming that the observed resonance pattern of the reactions continues in a similar way at lower energies, \citet{srr07} state that their rates at lower temperature represent the non-resonant, lower limits to the true values at $E \simeq 1.5\ \mathrm{MeV}$.

On the other hand, it is quite clear that the new rates are not relevant for the explosive nucleosynthesis of SNe Ia, which occurs in a different burning regime (nuclear statistical equilibrium) and at $T > 10^9\ \mathrm{K}$, where the corresponding experimental uncertainties on the astrophysical factor are much better constrained. 

The relevance of the new rate in the range $7 \times 10^8\ \mathrm{K} < T < 9.8 \times 10^8\ \mathrm{K}$ motivates therefore the investigation of its effect on the ignition process in SNe Ia, which typically occurs at these temperatures.

Due to the restrictions of the $\alpha$-network, in {\tt iso7} the C-burning is implemented in such a way that the reaction rate takes into account both the $\alpha$ and the proton channel, but the only reaction products are {\nuc 20 Ne} and $\alpha$. Since the $\alpha$ channel is dominant in the considered temperature range, this approximation is sensible.

It is worth mentioning that, besides the above cited measurements, in the last decade several works have been devoted to the theoretical prediction of the C-burning reaction rate. Some of them \citep{cll02,itw03} state that resonant screening effects could have a dramatic effect (either suppression or enhancement, respectively, for the two cited papers) on the reaction rate at low energy. \citet{jrb07} and \citet{gbc07} predict that the fusion cross section has a low-energy hindrance in a range of temperatures and densities which include the conditions in the WD core. \citet{gbc07} further compute the SN Ia ignition curve in this case, finding that, with the reduced burning rate, the ignition criteria will be fulfilled at a larger temperature. All these theoretical expectations will not be addressed explicitly in the present work.

\section{Results}
\label{results}

\subsection{Ignition in different WD models}
\label{results-models}

\begin{figure}
  \resizebox{\hsize}{!}{\includegraphics{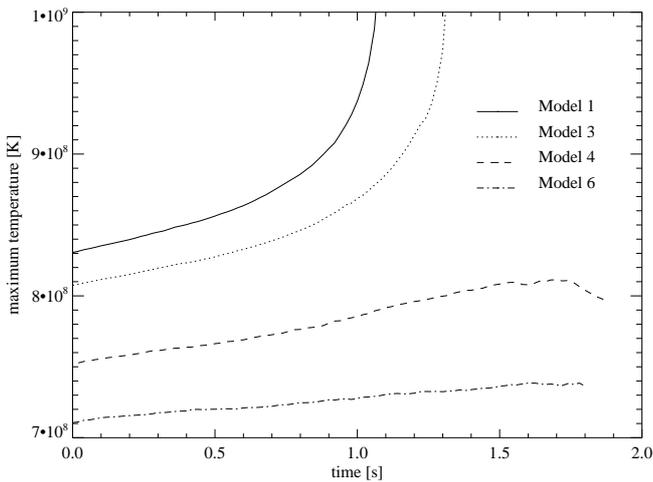}}
  \caption{Maximum bubble temperature as a function of time, in four bubble simulations with different WD models, reported in the legend. The initial bubble temperature is indicated at $t = 0$\ s, the other bubble parameters are set as described in Sect.~\ref{setup}.}
  \label{maxtemp}
\end{figure}

Four bubble simulations have been performed, one for each WD background model listed in Table \ref{models}. In these runs, the reaction rate is set as in \citet{cf88}. Figure \ref{maxtemp} shows the time evolution of the maximum bubble temperature in the four simulations. The bubble calculations with models 1 and 3 go to thermonuclear runaway, whereas in the cases 4 and 6 the bubble is fragmented by the hydrodynamical instabilities and its own rising motion before reaching the runaway. The typical evolutionary timescales are about 25\% longer than the ones reported in \citet{ibh06}, because of the lower $^{12}$C abundance in the present work.

Two main points explain the differences in the bubble simulations shown above. First, the initial bubble temperature determines the nuclear timescale. A colder temperature perturbation has a longer ignition timescale $\tau_{\mathrm{n}}$, according to 

\begin{equation}
\label{tau-nuc}
\tau_{\mathrm{n}} \propto T^{-22}\ \rho^{-4}
\end{equation}
\citep{wwk04,ibh06} and thus it is dispersed more efficiently by its own buoyant rise. 

Secondly, WD models with larger initial mass are colder and denser in the center, and have a larger gravitational acceleration at the initial bubble location (cf.~Table \ref{models}). This leads to a larger buoyant velocity $v_{\mathrm{b}}$ and thus to smaller dispersion timescales $\tau_{\mathrm{disp}}$ in models 4 and 6, which can be estimated as

\begin{equation}
\label{tau-rti}
\tau_{\mathrm{disp}} = \frac{D}{v_{\mathrm{b}}}
\end{equation}
where $D$ is the bubble diameter.

The ignition condition for a temperature perturbation is met when its nuclear timescale is smaller than the dispersion timescale. In the case of models 4 and 6, the bubble does not go to thermonuclear runaway: for this set of bubble parameters, the dispersion is faster than the nuclear burning. Suppose that we reduce the bubble distance from the WD center. At the same time the ignition timescale decreases (because the temperature increases) and the dispersion timescale increases (because $g$ decreases towards the center). Both effects make the ignition more likely.

The main result for this part of our study is therefore a dependence of the ignition features on the accreting history of the WD. WDs with a larger initial mass develop a core which is denser and cooler, according to \citet{lht06}. In these models we find that, for a given bubble size, the ignition is favored at a smaller distance from the WD center than in models with a smaller initial mass.  This interpretation is further strengthened by the computation of the burning zone radius $r_{\mathrm{b}}$ of the WD models (Table \ref{models}, fifth column), defined as the radius within which one-half of the luminosity is generated \citep{wwk04}. This radius can be considered a good approximation for the size of the region where the temperature perturbations arise. In denser models, $r_{\mathrm{b}}$ is smaller, with a maximum difference between models 1 and 6 of $25\%$. Temperature perturbations are thus likely to be generated closer to the WD center.

\subsection{The role of carbon burning}
\label{results-burning}

\begin{figure}
  \resizebox{\hsize}{!}{\includegraphics{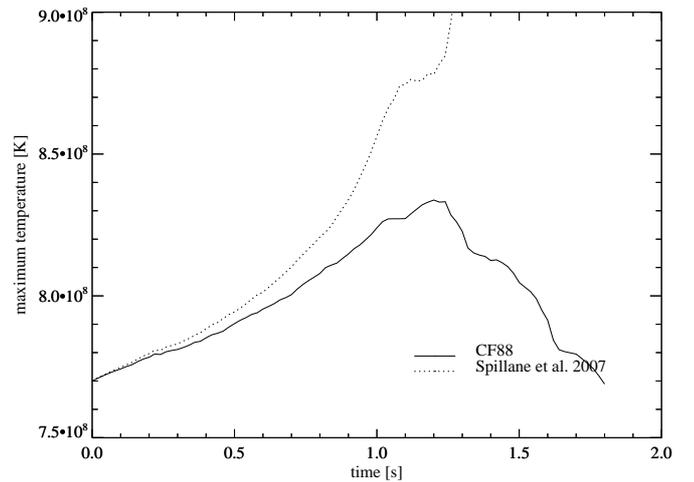}}
  \caption{Maximum bubble temperature as a function of time in two bubble simulations, with the same bubble parameters (described in the text) but a different prescription for the reaction rate of C-burning, as indicated in the legend.}
  \label{bubble-mod}
\end{figure}

The increase of the rate of C-burning in \citet{srr07}, though not dramatic, has a relevant impact on the evolution of temperature perturbations, due to the high power dependence on $T$ in Eq.~\ref{tau-nuc}. This is shown in Fig.~\ref{bubble-mod}, which refers to two simulations with the same bubble parameters but different references for the C-burning rate, namely \citet{cf88} and \citet{srr07}. The WD background model is the same as in \citet{ibh06} and the bubble parameters are $T = 7.7 \times 10^8\ \mathrm{K}$, $D = 1\ \mathrm{km}$ and $R = 100\ \mathrm{km}$. Not unexpectedly, the bubble simulation with the new reaction rate reaches the thermonuclear runaway, the other does not. From an analytical viewpoint, this is well described by the equation linking the reaction rate $\dot{S}$ and $\tau_{\mathrm{n}}$ \citep{wwk04}:

\begin{equation}
\label{nuclear-timescale}
\tau_\mathrm{n} = \left( \frac{1}{\dot{S}}\, \frac{d\dot{S}}{dt}
\right)^{-1} \approx \;  \left( \frac{1}{\dot{S}}\, \frac{\partial
\dot{S}}{\partial T}\, \frac{\partial T}{\partial t} \right)^{-1} = \;
\frac{C_\mathrm{p} T}{23\, \dot{S}} 
\end{equation}
where $C_\mathrm{p}$ is the specific heat at constant pressure. At $T = 8 \times 10^8\ \mathrm{K}$, $\dot{S}$ is 1.2 times larger, thus $\tau_{\mathrm{n, new}} \approx 0.8\ \tau_{\mathrm{n, old}}$, and a smaller nuclear timescale favors the bubble ignition before its fragmentation. 

This is not a surprising result, and has a limited impact on our knowledge of the ignition process. Another interesting question pertains to the effect of the new rates not only on a temperature perturbation, but on the overall WD structure before ignition. In other words, can a larger rate of C-burning modify the WD models before ignition \citep{lht06} and, consequently, the features of this process? 

\begin{figure}
  \resizebox{\hsize}{!}{\includegraphics[angle=90]{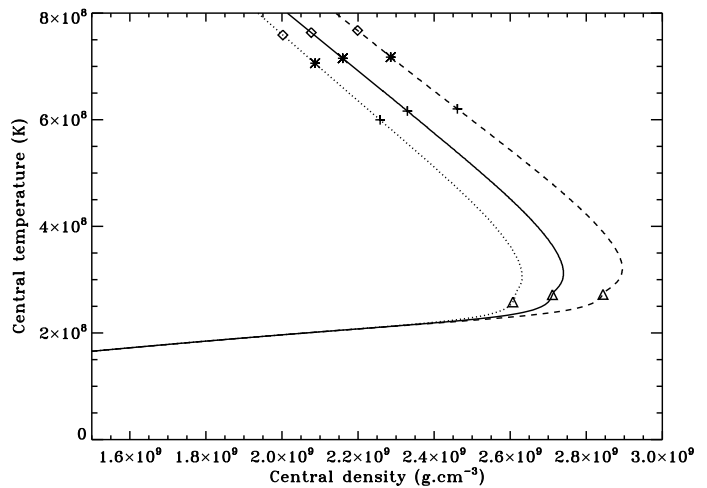}}
  \caption{Evolution for the central temperature and density of the
  star in Models 3 (solid), Model 3.1 (dashed) and Model 3.2
  (dotted). Symbols denote the birth of the convective core (triangles)
  and ignition points according to various criteria: the burning time
  scale is 1/22nd of (crosses) 1/8th of (stars) or equal to (diamonds)
  the convective turnover timescale \citep[see][]{lht06}. Although the
  position of the crosses and stars corresponds to actual output
  models of the stellar evolution code, the diamonds result from an
  extrapolation involving models at a lower temperature.}  
\label{luigi}
\end{figure}

In order to investigate this point, we run two additional WD models
with the code used in \citet{lht06} ({\small FLASH\_THE\_TORTOISE}\footnote{Code web page:  http://astro.ens.fr/?lesaffre.}) in the same
conditions as in Model 3, but we varied the adopted rate of
C-burning (Table \ref{models2}). Model 3.1 was run with the rate of \cite{srr07} instead of
the \citet{cf88} rate used for Model 3. The extrapolation of the rate at low temperature is about half the \citet{cf88} rate
in the low temperature range between $10^8\ \mathrm{K}$ and $5 \times 10^8\ \mathrm{K}$ (see
Fig.~\ref{rate-ratio}). As seen on Fig.~\ref{luigi}, the central part
of the star spends more time at low temperature in that case: as a
result the ignition curve is shifted to higher densities and the star
ignites at a slightly higher density (about 6\% higher, see Table
\ref{models2}).  The \citet{srr07} extrapolated burning rate is a lower estimate at low temperature, hence we can take the ignition density value of model 3.1 as
an upper estimate.  The gravitational acceleration follows the same trend as the central
density. On the contrary, the ignition temperature is almost
unchanged. This reflects the fact that the reaction rate is an
extremely steep function of the temperature. The burning zone radius
 is almost unchanged as well.

Model 3.2 was run with twice the \citet{cf88} rate, in order to check how big
changes in the rate, caused by possible discoveries of new low-lying resonances, can affect the ignition conditions (for example,
the \citealt{srr07} rate is, at most, only 20\% higher than the \citealt{cf88}
rate). Since the rate of burning is now everywhere higher than the
\citet{cf88} rate, the ignition curve on Fig.~\ref{luigi} is shifted to the
left hand side and the star ignites at lower density (by about 4\%, as
seen on Table \ref{models2}). As in model 3.1, the ignition
temperature is almost unchanged.

To conclude this brief investigation, the main effect of the rate is
rather surprising at low temperature, where it controls the density
at which the growth of the convective core starts, and hence
determines the future ignition density. The ignition
temperature is almost unaffected. The results on ignition temperatures and burning radii qualitatively agree with some calculations done with the heuristic convection model from \citet{ww04} and the reaction rates described above, for an isotropic flow and neglecting non-linear effects on the burning.

\begin{table}
\begin{minipage}[t]{\columnwidth}
\caption{Main features of the two additional WD models computed to investigate
the effects of changing the burning rate (quantities are same as in Table
\ref{models}).} 
\centering
\renewcommand{\footnoterule}{}  
\begin{tabular}{ccccc}
\hline \hline
Model & T$_\mathrm{c}$ [$10^8\ \mathrm{K}$] & $\rho_\mathrm{c}$
[$10^9\ \mathrm{g\ cm^{-3}}$] & $g$ [$10^9\ \mathrm{cm\ s^{-2}}$] &
$r_{\mathrm{b}}$ [km]\\
\hline
3.1 & 7.68 & 2.20 & 6.23 & 150\\
3 \footnote{As for model 3, the stellar evolution code
stalls shortly before the actual criterion for ignition is met, hence
these values result of an extrapolation from about 5\% lower
temperature models. However, we do not expect the extrapolation errors
to change much either the absolute or the relative values of the
numbers presented in this table.}   & 7.63 & 2.08 & 5.90 & 154\\
3.2 & 7.59 & 2.00 & 5.20 & 164\\
\hline
\end{tabular}
\label{models2}
\end{minipage}
\end{table}

\subsection{Numerical tests on the bubble simulations}
\label{numerical}

 The interpretation of the bubble simulations presented in this work cannot be separated from an evaluation of the numerical issues of the setup. As already discussed in \citet{ibh06}, the main features of the bubble physics are the fragmentation due to the buoyant motion, and (if the bubble does not reach the thermonuclear runaway) the subsequent dispersion by heat diffusion. The latter process is expected on length scales ($10 - 100\ \mathrm{cm}$) which are one to two orders of magnitude smaller than the effective spatial resolution of the performed simulations. Nevertheless, the physics which drives the dispersion, namely the interplay between nuclear burning and hydrodynamical instabilities, is correctly taken into account in the model, as the parameter studies in \citet{ibh06} show. Therefore one can reasonably expect that the numerical dispersion mimics, on larger length scales, the effect of the cooling by heat diffusion at unresolved scales. 

One can still argue about other oversimplifications in the bubble setup. For example, the choice of a spherical bubble shape can be debated. Numerical tests were performed with bubbles in stirred backgrounds, where the bubble is soon deformed and the growth of instabilities follows a different pattern with respect to the standard background model at rest \citep{ibh06}. The results show that, for well-posed choices of the stirring field, the turbulence-induced disruption (and thus the bubble shape) has not a leading effect in the bubble evolution.

Closely related to the bubble shape is the geometry choice: the 2D Cartesian geometry has been preferred to the more natural cylindrical, axisymmetric geometry for numerical reasons, after having verified that the results on bubble dispersion are very similar in both cases \citep{i05}.

\begin{figure}
  \resizebox{\hsize}{!}{\includegraphics{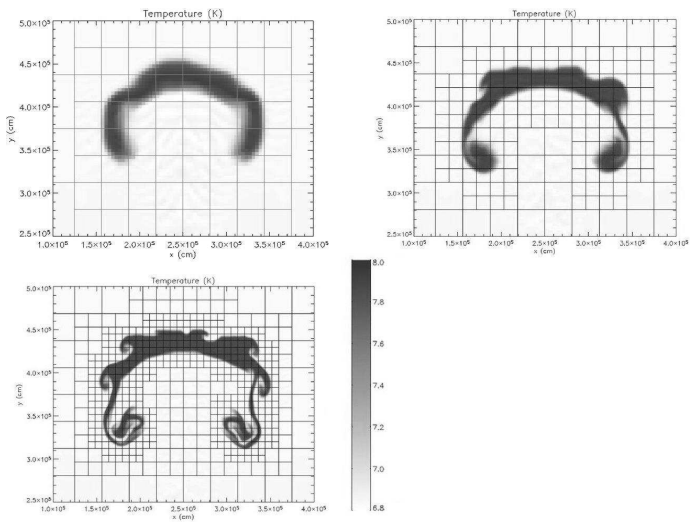}}
  \caption{Comparison of three temperature plots of buoyant bubbles,
after a simulation time of $0.5\ \mathrm{s}$, in runs with different maximum level of refinement. The plots enclose only a small
part of the computational domain ($3 \times 2.5\ \mathrm{km}$), 
centered on the bubble. The choice of the initial parameters is identical
in the three cases: temperature $7.7 \times 10^8\ \mathrm{K}$,
diameter $1\ \mathrm{km}$, distance from the center $100\ \mathrm{km}$, and the background WD model is the same as in \citet{ibh06}. Every mesh element in the plots is resolved in $8 \times 8$ grid cells. The color scale indicates temperature, in units of $10^8\ \mathrm{K}$. The panels refer to runs with four (upper left), five (upper right; this is the reference resolution of the runs presented in this work) and six (lower left) AMR levels.}
  \label{resolution}
\end{figure}

Another numerical issue concerns the discretization of the bubble on the Cartesian grid. This procedure may introduce errors, potentially able to seed instabilities that ultimately can dominate the evolution. A resolution study has been therefore performed \citep{ibh06} to verify whether the chosen level of resolution allows an adequate modelling of the hydrodynamical instabilities and the subsequent fragmentation. In Fig.~\ref{resolution} one can notice that different maximum levels of refinement resolve more and more Rayleigh-Taylor unstable modes. On the other hand, the resolution studies of the temperature evolution and the bubble dispersion \citep[Figs.~3 and 4 of][]{ibh06} clearly have convergent results for the two more resolved runs, thus indicating that the bubble evolution is dominated by large-scale instabilities, rather than by discretization errors.

\section{Discussion and conclusions}
\label{conclusions}

In this work we explored two previously unaccounted parameters of the ignition theory of SNe Ia, and studied how they can affect the features of the ignition process. The tools for this study are 2D numerical simulations of temperature perturbations in the WD core. In this framework,
the bubble model developed by \citet{ibh06} has several intrinsic and numerical limitations in its predictive power of the ignition properties of SNe Ia, but it can be profitably used as a ``probe'' of the influence on the ignition process of other physical parameters\footnote{For example, following the same approach, \citet{ibh07} propose that the WD composition affects also the ignition features (as suggested by \citealt{dt06}) and not only the SNe Ia explosion (cf.~\citealt{tbt03,rgr06,tjc09}).}.

Four WD models, derived by the stellar evolution and accretion history of the progenitors \citep{lht06}, were used as background for the bubble evolution. The thermonuclear runaway is not reached in the coolest and densest WD models 4 and 6, resulting from the evolution of initially more massive progenitors. We infer that, in these models, the ignition region lies closer to the WD center, where the gravitational acceleration is smaller and the bubble dispersion less effective. 

 The simplified bubble model that we used does not allow an accurate determination of the extent of the ignition zone.  An estimate of this quantity is provided by the size of the burning zone radius $r_{\mathrm{b}}$, where one-half of the nuclear energy of the WD is produced and where the temperature perturbations are likely to be generated. The burning zone radius is in the range between $170$ and $120\ \mathrm{km}$ and decreases in more massive progenitors, thus supporting the results of bubble simulations.

Clearly these estimates totally neglect any superimposed feature of the WD convection, such as a dipole. In large-scale simulations of the progenitor's hydrostatic evolution, on the other hand, the spatial resolution is insufficient to resolve the generation and ignition (or fragmentation) of the temperature perturbations. Therefore, we suggest to implement the key properties of these objects using a stochastic approach (cf.~\citealt{sn06}).

The second question addressed by this work is about the nuclear physics uncertainties in the ignition theory. Starting from the new measurement of \citet{srr07}, we study the impact of a different (larger) reaction rate for the C-burning on the ignition features in SNe Ia. This analysis is not limited to the temperature perturbations, but is extended to the WD structure by computing new stellar evolution tracks with various rates of C-burning. According to these models, the ignition features only have a mild dependence on the rate at low temperature (where lower rates favor higher ignition densities) and almost no dependence on the rate at high temperature. A somewhat similar analysis has been performed by \citet{csb09}: they study the possible effect of a hypothetical resonance at $E = 1.5\ \mathrm{MeV}$ by constructing ignition curves for the progenitor. They also find that such a feature would play no major role on the ignition process, though lower-lying resonances could. It is important to stress that the effect of different ignition conditions is manifold and not merely related to the location of the ignition zone; \citet{csb09}, for example, conjecture an impact on the iron-peak explosive nucleosynthesis \citep{pb08,cbt08} and on the velocity of the flame front \citep{rgr06}.

The whole field of the theoretical modeling of SNe Ia is reaching in recent years a mature stage, with growing interesting insights in the microphysics of these events (like for example the mechanism for DDT, \citealt{rn07,r07,w07}). Moreover, modern simulations provide a more and more detailed comparisons with the observables \citep{hsr07,mrb07,kw07}. The use of SNe Ia for precision cosmology has finally come on safer grounds, thanks to a deeper theoretical understanding of the involved physical parameters \citep{rhs07,krw09}. At the same time, novel low-Mach number schemes permit to fill the gap between the evolution of the progenitor and the onset of the explosion \citep{zab08,zab09}. In this context, the present study on the ignition parameters, although performed with very simplified tools, can drive further and more sophisticated analysis towards interesting issues, which call for a deeper investigation. 

\begin{acknowledgements}
Thanks to the referee for the constructive comments, which improved the presentation of this work.
The {\small FLASH} code is developed by the DOE-supported ASC / Alliance Center
for Astrophysical Thermonuclear Flashes at the University of
Chicago. L.I.~thanks the organizers of the {\it Catania Workshop on Nuclear and Neutrino Astrophysics} (WNNA 2007) for the fruitful meeting, which inspired part of this work. P.L.~thanks the French embassy in the United Kingdom for an overseas fellowship at Churchill College during which part of this work was conducted.
\end{acknowledgements}

\bibliography{snia-ref}
\bibliographystyle{bibtex/aa}

\end{document}